# Pointing Error Compensation For Inter-Satellite Communication Using Multi-Plane Light Conversion Spatial Demultiplexer

Antonin Billaud
*Cailabs*
*Rennes, France*
Antonin@Cailabs.com

David Allioux
*Cailabs*
*Rennes, France*
David@Cailabs.com

Nicolas Laurenchet
*Cailabs*
*Rennes, France*
Nicolas@Cailabs.com

Pu Jian
*Cailabs*
*Rennes, France*
Pu@Cailabs.com

Olivier Pinel
*Cailabs*
*Rennes, France*
Olivier@Cailabs.com

Guillaume Labroille
Cailabs
Rennes, France
Guillaume@Cailabs.com

*Abstract*— In this work, we study the effect of beam deviation angle at the reception side and calculate the theoretical demultiplexed collected energy for up to 15 modes, investigating the influence of the ratio between incoming beam size and fundamental HG mode's waist. We show this approach greatly enhances the collection efficiency, tolerating tip-tilt error of more than 3 times compared to a Gaussian beam alone, as shown in Figure 1. We also show that, depending on waist size, a trade-off between collection efficiency at small angles and maximum acceptance angle can be achieved.

## I. INTRODUCTION

Laser communication is a booming market for various applications due to the deployment of new technologies and the need for a worldwide access to data. This is particularly valid for non-terrestrial communications where free-space communications is the only option available. However the current RF technology is a limiting factor for high throughput achieving only Mbs/s whilst Gbs/s to Tbs/s throughput could be achieved via optical-based technology.

Satellite constellations is the next area where laser communication is to be deployed with thousands of satellites expected to be launched in the next few years. To enable intra-constellation link, laser communication is considered as one of the most efficient emerging technology. Compared to micro-wave, optical links can provide higher throughput with lighter, smaller and less consuming equipments [1].

Acquisition, tracking and pointing mechanisms are crucial to link's reliability as pointing errors result in signal degradations. To compensate it, different technologies can be deployed like gimbals, active systems or even adaptive optics on ground-based stations [2]. However, in satellite embedded technologies, available space, power consumption and environmental constrains limit the user to simpler solutions such as fast steering mirrors [3].

Here we propose a new passive architecture to compensate for pointing errors. This novel scheme utilizes a spatial demultiplexer based on the Multi-Plane Light Conversion (MPLC) technology [4]. Through a Taylor expansion, a Gaussian beam misalignment can be converted to a sum of low order Hermite-Gaussian (HG) modes [5]. Using a spatial demultiplexer at the optical links' reception side, it is thus possible to collect misaligned beams. The MPLC passively converts HG modes to the same number of beams coupled to SMF that can benefit from mature telecom technology.

## II. MULTI-PLANE LIGHT CONVERSION

MPLC is a technique that allows performing any unitary spatial transform. Theoretically, it enables lossless conversion of any set of $N$ orthogonal spatial modes into any other set of $N$ orthogonal modes. Conversion is done through a succession of transverse phase profiles each separated by a free-space propagation serving as a fractional Fourier transform operation. Principle of the MPLC is shown schematically in Figure 1.

In particular, MPLC enables mode selective spatial multiplexing and demultiplexing, i.e. the conversion of $N$ spatially separate input Gaussian beams into N orthogonal modes. Practically, MPLC is implemented using a multi-pass cavity, in which the successive phase profiles are all manufactured on a single reflective phase plate. Figure 2 shows a picture of a typical MPLC. Cavity is formed by a mirror and the reflective phase plate, noticeable with its gold coating on the picture. An MPLC used in the reverse direction implements the demultiplexing operation.

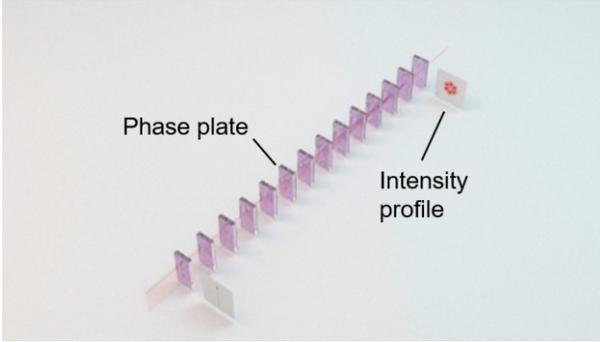

Figure 1: Multi-Plane Light Conversion principle.

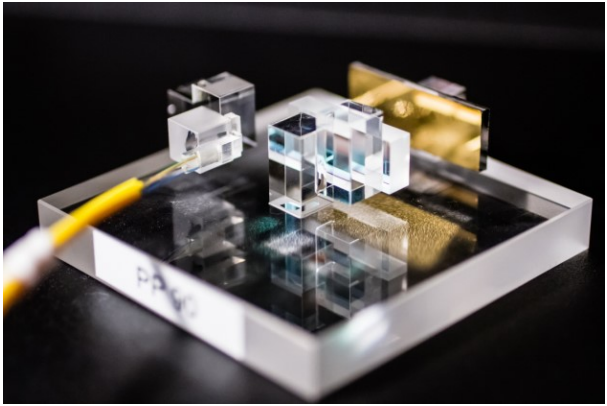

Figure 2: Photo of an MPLC. On the left, the input consists of an array of single mode fibers. The phase plate is coated in gold. In front of it, a mirror operates as the second element of the cavity. The free space output exits on the right side of the picture.

### III. ONE DIMENSION DEPOINTING

In this study, we consider a beam sent from one satellite to another from a large distance. When the beam arrives on the receiver terminal, it can be considered as a top-hat beam with a flat phase. Considering that an apodization occurs through the satellite aperture, we approximate the top-hat by a super-gaussian of order 3. We consider the input beam with a 12.5 mm waist (in $1/e^2$), corresponding to a 25 mm aperture. Depending on the depointing occurring in this data link, a linear phase can be observed on the received beam as shown in Figure 3.

We investigate the pointing error compensation via a MPLC with a set of Hermite-Gaussian (HG) modes. As described in reference [5] the HG basis is convenient for pointing error compensation as a small displacement corresponds to an apparition of higher-order modes through a Taylor expansion. In other words, a displacement naturally translates into Hemite-Gaussian modes. Moreover, it is a convenient basis for the MPLC as it allows to increase the number of modes while keeping similar performances.

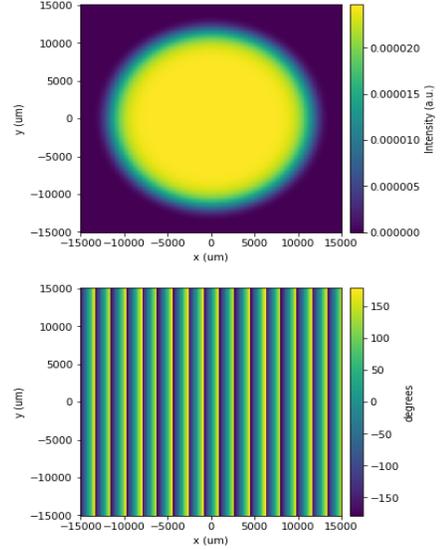

Figure 3: Example of intensity (up) and phase (down) profile of the incoming beam.

We start by investigating the collected light in one dimension. The multiplexer main two parameters that we can play with are the number of modes and the waist of the modes. Figure 4 to Figure 7 shows the average collection efficiency as function of the maximum depointing for 3 to 45 Hermite-Gaussian modes with a waist from 2.5 mm, 3.75 mm, 5 mm and 7.5 mm.

As expected, the coupling efficiency increases with the number of modes. More interesting is the influence of the waist on the coupling efficiency. While, as a first guess, we were expecting an optimum waist value, the MPLC acts as a low-pass filter on the depointing angle with the cutoff angle and maximal collection depending on the waist.

For a small waist of 2.5 mm for example, we observe that the collection efficiency is low, with a maximum of 50% for 45 modes and 30% for 15 modes, but the slope is quite small allowing us to collect more than 25% and 15% respectively for 70 arcsec. The collection is weak but only a factor of two is measurable between low and high angle.

When the waist increases, improvement of the collection efficiency is observed, particularly for low depointing angles with a more drastic low-pass filter behavior. Maximum collection efficiency is demonstrated for a 7.5 mm waist where we measure >90% efficiency until 40 arcsec but reaches 0% efficiency at 70 arcsec. Figure 8 sums up different behavior for a 15 modes multiplexer that is good compromise between performances and number of modes.

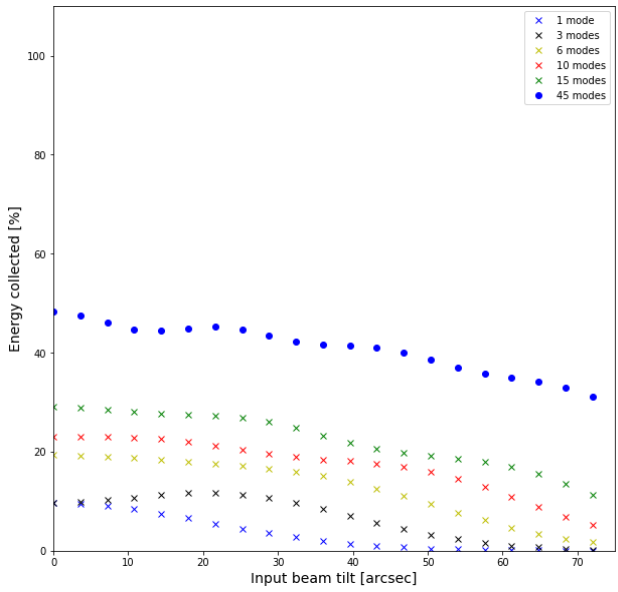

*Figure 4: Average collection efficiency as a function of the input angle for an MPLC mode waist of 2.5 mm*

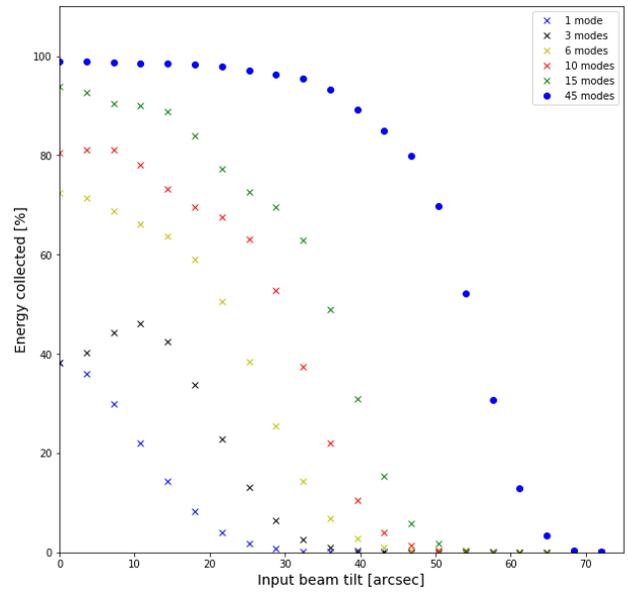

*Figure 6: Average collection efficiency as a function of the input angle for an MPLC mode waist of 5.0 mm*

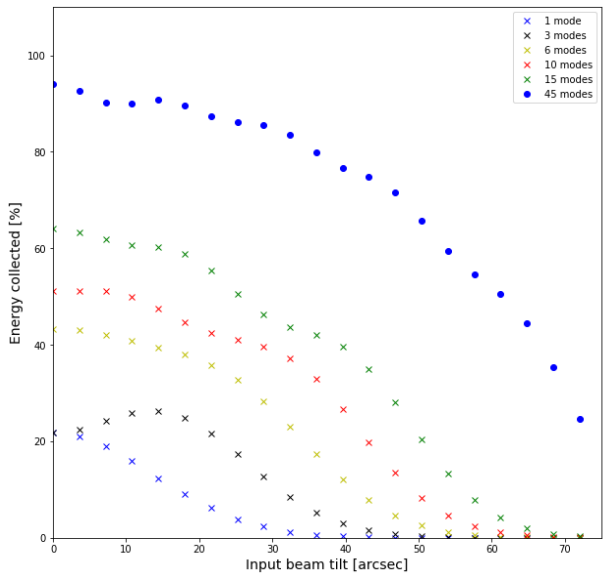

*Figure 5: Average collection efficiency as a function of the input angle for an MPLC mode waist of 3.75 mm*

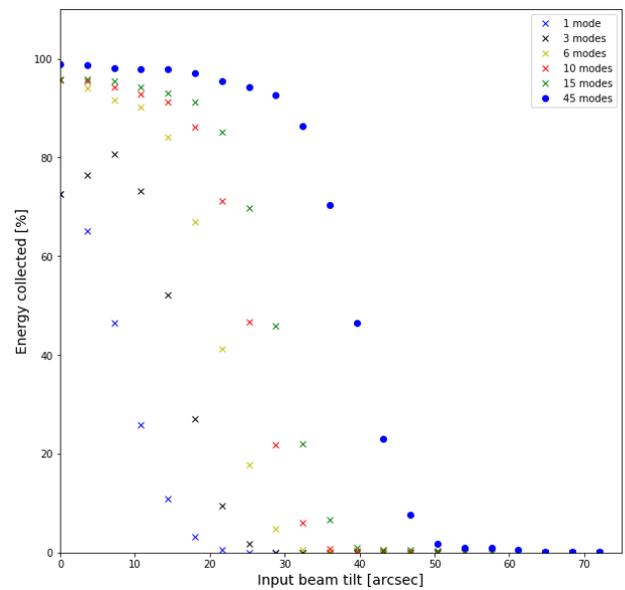

*Figure 7: Average collection efficiency as a function of the input angle for an MPLC mode waist of 7.5 mm*

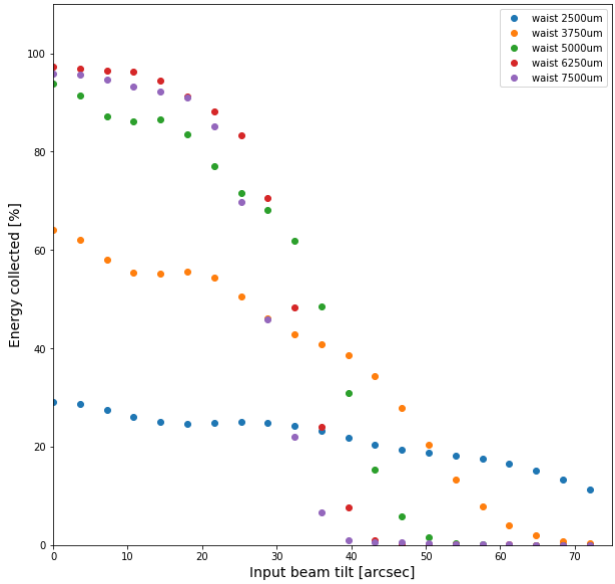

*Figure 8: Average collection efficiency as a function of the input angle for a 15 modes multiplexer*

To illustrate this possible trade-off, we calculate the collection efficiency for an extreme case of up to 72 arcsec for an increasing number of HG modes for two waists of 3 mm and 5 mm. The results are plotted respectively in Figure 9 and Figure 10. The number of modes is in the x-axis and the collection efficiency in the y-axis. Each point corresponds to the efficiency for a given collection angle and number of modes. For the smaller waist (Figure 9), maximum efficiency is weak, but at the same time, as soon as we reach a minimum number of modes, approximately 15, we always collect a portion of the energy for the all the depointing angles. With 45 modes, the collection is confined between 30% and 70%.

On the other side, for a larger waist (Figure 10), maximum collection efficiency increases very fast. With only 6 modes, some of the angles are collected with more than 70%. However, some depointing angles are never collected and this, even with 45 modes.

## IV. Conclusion

In this work, we studied the MPLC technology approach to compensate for pointing errors. We investigated the collection efficiency for different angles and different number of modes in the MPLC. We showed this approach greatly enhances the collection efficiency compared to a single-mode fiber setup. Depending on the MPLC modes' waist size, a trade-off between collection and maximum acceptance angle can be achieved. We believe this passive approach is well suited to small high speed depointings compensation.

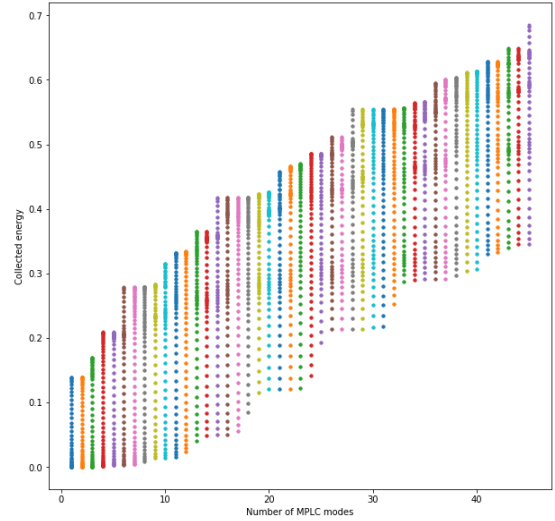

*Figure 9: Collection efficiency as function of the number of modes for all the collection angles for 3 mm fundamental mode waist*

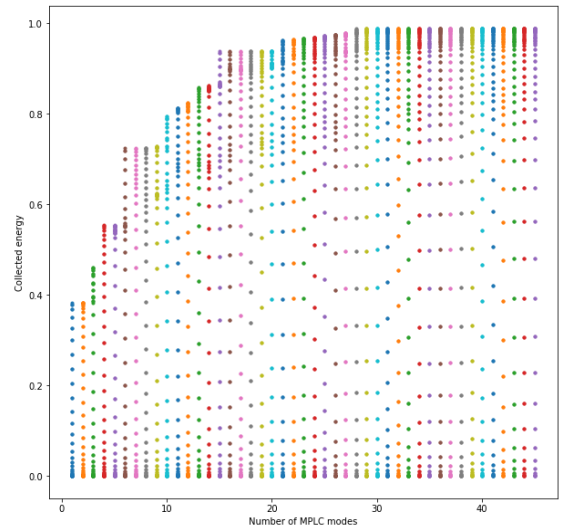

*Figure 10: Collection efficiency as function of the number of modes for all the collection angles for 5 mm fundamental mode waist*